\documentclass[psfig]{aa}
\input psfig.sty

\begin{document}

   \thesaurus{06     
              (08.04.1;  
               08.05.3;  
               08.06.3;  
               08.08.1;  
               08.16.5;  
               08.22.2)} 

   \titlerunning{Pulsation in Herbig Ae stars}
   \title{Pulsation in two Herbig Ae stars: HD~35929 and 
	  V351~Ori\thanks {Based on observations carried
          out  at the European Southern Observatory, La Silla, Chile under 
          proposals number 62-I-0533, 63-I-0053}}

   \author{ 
   M. Marconi\inst{1},
   V. Ripepi\inst{1},
   J.M. Alcal\'a\inst{1},
   E. Covino\inst{1},
   F. Palla\inst{2},
   \and
   L. Terranegra\inst{1}
          }

   \offprints{V. Ripepi: email: ripepi@na.astro.it}

   \institute{Osservatorio Astronomico di Capodimonte, Via Moiariello 16
               I-80131 Napoli, Italy \\
              Osservatorio Astrofisico di Arcetri, Largo E. Fermi 5, I-50125 
               Firenze, Italy \\
}

\date{Received/Accepted}

\maketitle

\begin{abstract}
New photometric observations of seven intermediate mass pre-main sequence
$\delta$ Scuti candidates are presented. 
The periods and pulsation modes are derived for two of these  
stars, namely HD~35929 and V351~Ori. The comparison between 
observations and nonlinear pulsational 
models allows us to provide some initial constraints on their mass and 
evolutionary state. As an illustration we discuss 
the use of periods to identify the mode of pulsation 
in these two stars and to have an independent estimate of their distances.

      \keywords{Stars: distances -- 
                Stars: evolution -- 
                Stars: fundamental parameters --
                Stars: HR diagram  -- 
                Stars: pre-main sequence -- 
                Stars: delta scuti
               }

\end{abstract}

%

\section{Introduction}

Young pre--main-sequence (PMS) stars of mass greater than $\sim$1.5 M$_\odot$
cross the region of pulsational instability during their contraction toward 
the main-sequence. The location of the instability strip
in the H-R diagram has been recently identified
by means of nonlinear models for the first three radial modes (Marconi
\& Palla 1998). The time spent by intermediate-mass stars within the boundaries
of the strip represents a small fraction of the
Kelvin-Helmoltz time scale, varying between $\sim$10\% for a star of 
1.5 M$_\odot$ and $\sim$5\% for a 4 M$_\odot$ star. Despite the brevity
of this phase, however, a number of known Herbig Ae stars have the
appropriate combination of luminosity and effective temperature
to become pulsationally unstable. In our previous study, we suggested
to look for $\delta$~Scuti-type photometric variations with periods of
minutes to several hours
and amplitudes less than few tenths of magnitudes in a sample
of Herbig Ae stars whose position in the H-R diagram coincides with
the instability strip. \\
The identification of a few PMS objects pulsating as
$\delta$ Scuti stars (Breger 1972), the prototype being the 
star HR~5999 (Kurtz \& Marang 1995), has provided some support to the
connection between variability and stellar pulsation. 
Of particular interest is the Ae star HD~104237 
that shows both short- and long-term velocity changes of
spectral lines (Donati et al. 1997). These variations indicate that
the star is undergoing radial pulsations with a period of approximately
40 minutes and an amplitude of about 1 km s$^{-1}$ (see also B\"ohm et al.
2000, in preparation). Interestingly, HD~104237
is the first intermediate mass PMS star with a measured magnetic field
(Donati et al. 1997).
A few new PMS candidate pulsators have been recently identified
by Pigulski et al. (2000).\\
Stimulated by these initial results, we have started a photometric
investigation of a sample of seven Herbig Ae stars with spectral types in
the range A5 to F5, located within or near the boundaries of the
instability strip.
For some of them, large time scale variations
have been observed during the long term monitoring program of
variable stars conducted at ESO (LTPV project: Sterken et al. 1995 and 
references therein).
However, no information is available on their variability 
on time scales shorter than 2 or 3 days. \\
The main goal of our study is to detect and characterize the
pulsation properties of young stars.
This way, we can improve our knowledge of their internal
structure and obtain unique constraints on the theoretical
predictions of the models. Ultimately, the analysis of the pulsation
characteristics
can yield an indirect estimate of the stellar mass. This represents
a powerful method for stars that are not part of the restricted group
of spectroscopic binary systems.
In this Letter, we report the discovery of two additional Herbig Ae stars 
which show evidence for variability of the $\delta$ Scuti-type.


\section{Selection of the sample, observations and data reduction}

The selection of the stars was based on their spectral type. As an initial
choice, we have
adopted values published in the literature. 
To have an independent check on the effective temperature 
of the selected stars, we have
used the Str\"omgren photometry provided by 
the ESO catalogues of the LTPV project. Then, we have placed the stars in the 
dereddened color-color [m1]--[c1] diagram (see Str\"omgren 1966), 
and compared their position with the theoretical colors 
from model atmospheres (Kurucz 1992). 
As a result, we derived a sample 
of seven stars with spectral types between F5 and A5: V346~Ori,
V351~Ori, BF~Ori, BN~Ori, NX~Pup, HD~35929, and AK~Sco.

 \begin{table}[h]
 \caption{Herbig Ae stars with $\delta$~Scuti type variability}
 \begin{tabular}{lcccc}
 \hline
 \hline
Star  & V-mag & SpT$^a$     & d [pc]$^a$ & P [d] \\ \hline\noalign{\smallskip}
 V351 Ori    &  8.9  & A7IIIe    & $>$210   & 0.058$\pm$0.001$^b$  \\
 HD 35929    &  8.2  & F0IIIe    & $>$360   & 0.196$\pm$0.005 \\
\noalign{\smallskip}
\hline
\hline
\end{tabular}

\footnotesize{a) van den Acker et al. (1998)};
\footnotesize{b) main pulsation period }
 \end{table}

The photometric observations were performed in two 
runs with different telescopes at ESO, La Silla. 
In the first period (Dec. 23--29, 1998),
we used the 0.5k$\times$0.5k CCD (Tektronix TK512CB) attached to the 
0.9m DUTCH telescope.
Five of the seven nights were photometric. 
In the second period (June 30 to July 4, 1999) we used the 2k$\times$2k CCD
(LORAL/LESSER) attached to the 1.5 m. DANISH telescope.
Unfortunately, weather conditions were excellent for accurate
photometric observations only during one night.
Due to their brightness, all of the sources are suitable for observations
in the Str\"omgren $uvby$-system. 
Two comparisons stars were used for each candidate variable; 
whenever possible, we observed the same comparison stars as in 
the LTPV program.
Data reduction and analysis to derive instrumental aperture magnitudes
were performed following the standard procedures under the MIDAS environment. 
The typical intrinsic instrumental photometric error in each Str\"omgren 
filter is of the order of a few  thousands of a magnitude.

\section{Results: $\delta$ Scuti candidates}

\begin{figure}[t]
 \psfig{figure=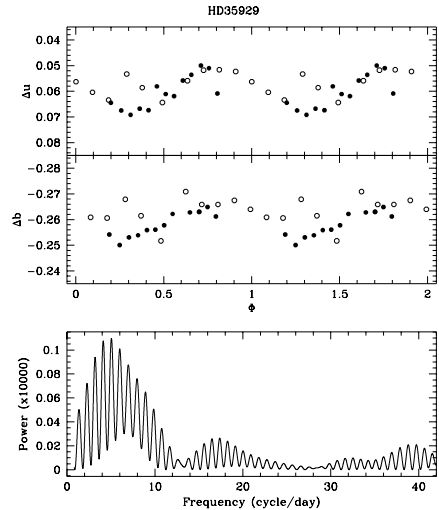,height=7.0cm}
 \caption{ {\bf Upper panels}: differential (variable$-$comparison star) 
 light curves of HD~35929 in the $u$- and $b$-filters. Filled and open 
 circles refer to the first and second night of observation respectively.
 {\bf Lower panel}: frequency spectrum of HD~35929. 
 \label{hd}}
 \end{figure}

The stars V346~Ori, V351~Ori, BF~Ori, HD~35929, and AK~Sco have
 been found to show photometric variability. For V351~Ori and HD~35929,
 we could derive reliable pulsational periods that fall in the expected 
range of $\delta$ Scuti type variability\footnote{As for V346 Ori and AK 
Sco, 
the variation seems $\delta$ Scuti-like, with $u$ amplitudes of about 0.035 
and 0.08 mag respectively but no period could be derived due to the 
poor phase coverage. As for BF Ori, its variation is clear 
($\Delta u$ $\sim$ 0.11 mag), but seems rather monotonic}. 
In Table~1 we report their main properties. \\
HD~35929 was observed during two nights (Dec. 28 and 29, 1998).
The stars HD~37399 and HD~37210 were used for comparison.
Unfortunately, HD~37210 turned out to be variable with a 
maximum amplitude in the $u$-filter of $\sim$0.03 mag and a rather 
monotonic variation. 
Therefore, the differential light curve of HD~35929 was derived 
using the data points relative to HD~37399 only. 
The instrumental phased light curves in $\Delta u$ and $\Delta b$
are shown in the two upper panels of Fig.~\ref{hd}.
The frequency spectrum of HD~35929 obtained
from the $u$-filter is shown in the lower panel of Fig.~\ref{hd}.
The frequency spectrum peaks at 5.1 d$^{-1}$, i.e. 0.196d. We note 
that no significant differences in the calculated period are found 
using light curves in other filters.\\

\begin{figure}[t]
 \psfig{figure=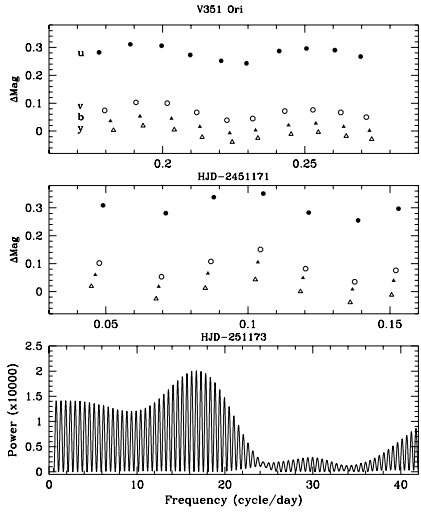,height=7.0cm}
\caption{{\bf Upper panels}: differential (variable$-$comparison star) 
light curves of V351 Ori in the labeled filters for the nights of  
Dec. 23 ({\bf top panel}) and Dec. 25 ({\bf middle panel}). 
{\bf Lower panel}: frequency spectrum of V351~Ori in the first
and second night of observation (solid and dotted lines, respectively).
The peak frequency (in days$^{-1}$) for each night is also indicated.
\label{v351}}
\end{figure}

V351 Ori was observed during the nights of Dec. 23 and 25, 1998. 
HD~38155 and HD~35298 were used as comparison stars. 
The latter turned out to be a variable star, with a maximum
amplitude in $u$ of $\sim$0.04 mag and a quasi-periodic behaviour.
Thus, we use only HD~38155 for comparison.
The differential magnitudes in each filter 
($\Delta u, \Delta v, \Delta b, \Delta y$, computed as variable$-$comparison) 
are shown in the upper panels of Fig.~\ref{v351}.
The light curve in the $u$-band, where the largest amplitude is 
observed, was used to determine the observed period of 
V351~Ori.
The frequency spectrum of V351 Ori obtained from the $u$-filter is 
shown in the lower panel of Fig.~\ref{v351}. The main maximum in this 
spectrum corresponds to a period of 0.058 d. Again, no significant differences
in the derived periods occur when using light curves in other filters.

\section{Evolutionary state of HD~35929 and V351~Ori}

\begin{figure}[t]
 \psfig{figure=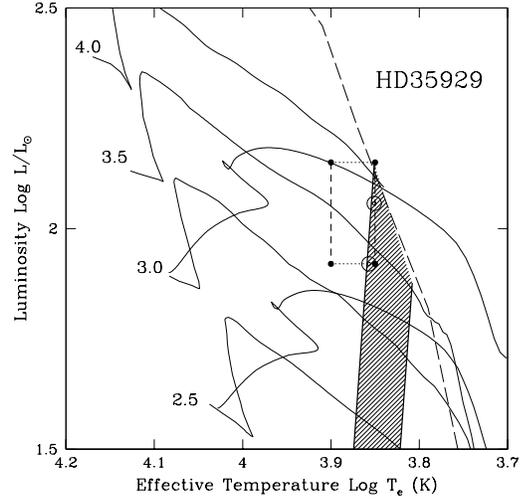,height=7.0cm}
 \caption{The position of HD~35929 in the H-R diagram according to the
  estimates of spectral type and distance found in the literature 
  (dotted box). The shaded region is the instability strip predicted 
  by Marconi \& Palla (1998). 
  The PMS and post-MS evolutionary tracks are shown as thick 
  and thin solid lines, respectively. The birth-line for the
  PMS evolutionary tracks is displayed by the dashed line.
 \label{hr_hd}}
 \end{figure}

Although more systematic observations are needed to define
the precise pulsational periods of HD~35929 and V351~Ori,
the available data
already provide some preliminary constraints on their
position in the H-R diagram, stellar mass and 
evolutionary state. As remarked by the referee, the discussion presented
in this section  is useful as an {\it illustration} of how well-defined
periods for these stars could indeed tightly constrain their 
evolutionary states.\\
Fig.~\ref{hr_hd} shows the location of HD~35929 in the H-R diagram. 
The dotted box accounts for the uncertainty in the spectral type 
(A5 to F0: Malfait et al. 1998, Miroshnichenko et al. 1997,
van den Ancker et al. 1998) and distance ($d=$360 to 430 pc: 
van den Ancker et al. 1998).
The instability strip for the first three radial modes, as predicted 
by Marconi \& Palla (1998), is also shown together with the PMS evolutionary 
tracks computed by Palla \& Stahler (1993) and
the post-MS evolutionary tracks 
for 2.5 and 3.0 $M_{\odot}$ of Castellani et al. (1999). 
The two circles indicate the best combination of the stellar 
parameters ($M$, $L$, $T_{\rm eff}$) that yield a period equal to the
observed one, $P=0.196\pm0.005$~d. 
These values are listed in Table~2.
The two solutions indicate a mass of 3.4 or 3.8~M$_{\odot}$, pulsating 
in the first overtone (FO) and second overtone (SO) respectively:
in both cases, HD~35929 can be considered a PMS pulsator, as expected. 

 \begin{figure}[t]
 \psfig{figure=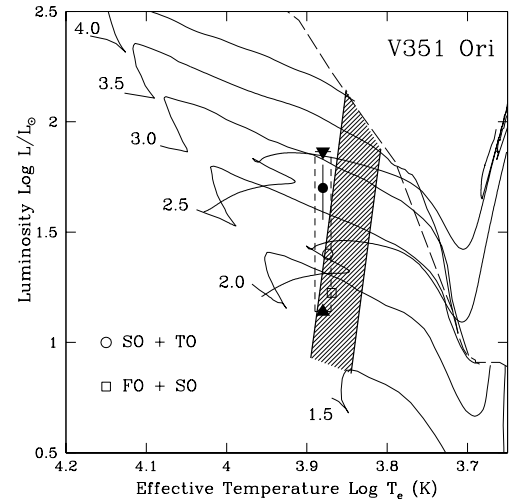,height=7.0cm}
 \caption{Same as Fig.~\ref{hr_hd} but for V351~Ori. The 
  filled triangles refer to the upper and lower limits of the 
  distance.
 The open symbols are the best pulsation models reproducing 
 an oscillation consistent with the observed period.
 \label{hr_v351}}
 \end{figure}

A combination of parameters for a post-MS stellar mass can also 
reproduce the pulsational period observed in HD~35929. In this 
case, the best choice would be a post-MS model of a 2.7~$M_{\odot}$ 
star with $L=83~L_{\odot}$ and $T_{\rm eff}=6900$~K, pulsating in the SO 
 mode. The location of this solution is quite close to the 
lower circle shown in Fig.~\ref{hr_hd}. Although only few specific studies
exist on HD~35929, some evidence supports the fact that HD~35929 
is a young star 
associated with the Ori OB-1c association. For example, 
Malfait et al. (1998) have discussed the infrared excess observed
toward this star, whereas Miroshnichenko et al. (1997)
have suggested that the star might be in a transition phase between
a PMS Herbig Ae star and a $\beta$ Pictoris-type object.
We also note that the pulsational character (period and H-R diagram location)
of HD~35929 is similar in 
many respects to 
that of the well known Herbig Ae star HR~5999.
From these considerations, we favor the conclusion that HD~35929 is a 
PMS star with a mass in the narrow
range 3.4-3.8~$M_{\odot}$, pulsating in the FO or SO.\\
Finally we note that if the Str\"omgren indices [c1] and [m1] 
measured for this star are taken
into account, together with published values for the $H\beta$ index,
one derives an effective temperature $\log{T_{\rm eff}=3.84}$ and a luminosity 
varying between 26 and 36 $L_{\odot}$. This means that according to 
the Str\"omgren
photometry, HD35929 should be located on a $\sim{2.5}M_{\odot}$ evolutionary
track and the period based on present data would be too long even for a 
pulsation in the fundamental mode. A possible explanation for this 
inconsistency could be that HD35929 is also a rapid rotator 
($\sim$150 Km s$^{-1}$) so that
the assumption of radial pulsation could be not completely correct.
Future observations and numerical simulations are needed in order
to properly address this problem.

 \begin{table}[h]
 \caption{Derived stellar parameters}
 \begin{tabular}{lcrccr}
 \hline
 \hline
Star  & Mass & $L$ &  $T_{\rm eff}$  & d & mode \\ 
   & ($M_\odot$) & ($L_\odot$) &  (K)  & (pc) & \\ \hline\noalign{\smallskip}
 HD~35929    &  3.4  & 83  & 7190  & 360 & FO  \\
             &  3.8  & 114 & 7100  & 420 & SO \\
             &  2.7$^a$  & 83  & 6900  & 360 & SO     \\
 & & & & & \\
 V351~Ori    &  1.85 & 17  & 7400  & 230 & SO  \\
             &  2.15 & 25  & 7480  & 280 & TO \\
 \noalign{\smallskip}
 \hline
 \hline
 \end{tabular}

\footnotesize{a) post-MS model}
 \end{table}

Fig.~\ref{hr_v351} shows the location of V351~Ori in the H-R diagram.
Here, the uncertainty on the luminosity is larger than for HD~35929, because
of the distance ambiguity. 
The lower value corresponds to the minimum distance of 260 pc 
given in the Hipparcos catalogue 
(van den Ancker et al. 1998). The upper 
limit (inverted triangle) assumes that V351~Ori is located in the Orion
molecular cloud at a distance of 460~pc. The dashed box corresponds to an 
uncertainty of $\pm0.01$ dex in $\log T_{\rm eff}$. Finally, the 
filled circle marks the position estimated by van den Ancker et al. (1996)
with the associated error bar. 
As already pointed out, present data for V351~Ori suggest a pulsation
 period of $0.058\pm0.002$. \\
Using the constraints provided by this preliminary period and by the
topology of the instability strip, 
we have computed linear nonadiabatic models to find the best
set of stellar parameters that reproduce the pulsation of V351~Ori.
The results are shown in Fig.~\ref{hr_v351} and the stellar parameters
are given in Table~2. The solutions yield a stellar mass of 1.85 M$_\odot$
or 2.15 M$_\odot$, respectively, pulsating in the SO 
(open square in Fig.~\ref{hr_v351}), or the third overtone 
(TO) mode (open circle in Fig.~\ref{hr_v351}. 
For lower modes, the luminosity of the model would be lower than the
estimated lower limit for V351~Ori, whereas higher modes are probably
excluded by the closeness of the observational box to the second
overtone blue boundary and we did not consider their occurrence. 
These solutions would tend to favor a distance of V351~Ori,
smaller than that of the young stellar population of the Orion complex. \\
Recently, Koval'chuk \& Pugach (1998) have argued on the basis of
several peculiar photospheric properties that V351~Ori
is in fact more evolved than previously thought and conclude that
this star does not belong to the group of Herbig stars.
From Fig.~\ref{hr_v351}, we see that the the post-MS track
of a 2 M$_\odot$ star intersects the corresponding PMS track at 
about the position of the best pulsational models. The degeneracy of 
the tracks does not allow to use the pulsational analysis to
discriminate between the two evolutionary phases. However,
this uncertainty would disappear if the distance of V351~Ori were 
the same as that of the Orion population stars.
Then, the difference between the pre- and post-MS tracks would be large
enough that the period analysis would rule out one of the two
solutions. Future studies of V351~Ori should address this important
aspect. \\
As in the case of HD35929, we used the measured Str\"omgren indices 
[c1] and [m1] to provide an independent evaluation of the location
in the HR diagram of this star. The results, ${\log{T_{\rm eff}}=3.89}$, 
$L\simeq 30L_{\odot}$, now suggest 
a stellar mass of $\sim 2.25 M_{\odot}$ for this pulsator, in which case 
the observed periods would be consistent with an oscillation in the 
 TO pulsation mode.\\
In conclusion, the present observations yield compelling evidence
for the occurrence of $\delta$ Scuti-type pulsation
in two Herbig Ae stars, HD~35929 and V351~Ori. The comparison with
evolutionary and pulsational models provides independent, even if
quite preliminary, constraints 
on the mass and evolutionary state of these stars.


\acknowledgements{We thank the referee, J. Matthews, for his very 
helpful report. We also thank Dr. R. Silvotti for useful discussions.
This research has made use of the Simbad database, operated by
CDS, Strasbourg, France.}


\end{document}